\documentclass[conference]{IEEEtran}
\IEEEoverridecommandlockouts    % to override locked commands

\usepackage{multirow}
\usepackage{lipsum}
\usepackage{amsmath}
\usepackage{amssymb}
\usepackage[ruled,vlined]{algorithm2e}
\usepackage{graphicx}
\usepackage{booktabs}
\usepackage{tabularx}
\usepackage{float}
\usepackage{xcolor}
\graphicspath{{./Figures/}}
\usepackage[pdftex,colorlinks=true,linkcolor=black,urlcolor=blue,citecolor=black]{hyperref}
\usepackage{xcolor}

\title{	\vspace*{18pt} \LARGE \bf Dynamic Gradient-Based Calibration\\for Robust and Accurate Traffic Macrosimulation}

\author{
	\parbox{\textwidth}{%
		\centering
		Shreyaa Raghavan$^{*}$, Cameron Hickert$^{*}$, Monica Chan, Cathy Wu%
	}%
	\thanks{$^{*}$These authors contributed equally to this work.}%
    \thanks{Shreyaa Raghavan and Cameron Hickert are with the Laboratory for Information \& Decision Systems and the Institute for Data, Systems, and Society, Massachusetts Institute of Technology, Cambridge, MA 02139, USA {(Email: \tt\small shreyaar@mit.edu; chickert@mit.edu})}%
    \thanks{Monica Chan is with the Dept. of Electrical Engineering and Computer Science, Massachusetts Institute of Technology, Cambridge, MA 02139, USA {(Email: \tt\small mochan@mit.edu})}%
    \thanks{Cathy Wu is with the Laboratory for Information \& Decision Systems, the Institute for Data, Systems, and Society, and the Dept. of Civil and Environmental Engineering, Massachusetts Institute of Technology, Cambridge, MA 02139, USA {(Email: \tt\small cathywu@mit.edu})}%
}%

\begin{document}
	
	\maketitle
	\thispagestyle{empty}
	\pagestyle{empty}
	
	%%%%%%%%%%%%%%%%%%%%%%%%%%%%%%%%%%%%%%%%%%%%%%%%%%%%%%%%%%%%%%%%%%
	\begin{abstract}

        Robust and accurate calibration of macroscopic traffic flow models such as METANET is critical for reliable prediction and effective control. While gradient-based methods are desirable for high-dimensional parameter spaces, their application to real-world traffic scenarios is hindered by highly nonconvex optimization landscapes. Consequently, standard static calibration frequently yields parameter sets that produce unstable, unrealistic traffic dynamics, undermining confidence in the estimated parameters and compromising the simulation's utility for counterfactual scenario testing. To address this, we propose a dynamic, rolling-horizon calibration framework. By reformulating static one-time estimation into a dynamic control problem, parameters better maintain stability and accuracy amid measurement noise. Using real-world data from the I-24 MOTION testbed, this work empirically characterizes the instability of standard methods. It then shows that the proposed approach simultaneously enhances robustness to perturbations and achieves a 48\% improvement in predictive accuracy over conventional static calibration.
	\end{abstract}
	
	%%%%%%%%%%%%%%%%%%%%%%%%%%%%%%%%%%%%%%%%%%%%%%%%%%%%%%%%%%%%%%%%%%
	\section{Introduction}
	\label{sec:introduction}
	% \begin{itemize}
 %            \item Traffic simulation is the backbone of ITS and control of traffic system. In particular, macroscopic simulation (define?) are valuable due their fast computation time. Various studies use them (cite), the most popular one being METANET.
 %            \item However, standard gradient-based methods of calibration often end up with a set of parameters that result in an unstable and unrealistic sim. This is in large part due to the extremely nonconvex and nonlinear nature of the parameter landscape, and as a result, macrosim ends up in local minima. Additionally, du, tiny perturbations or noise can destabilize traffic, which makes the simulation useless in counterfactual scenarios. While some literature explores this, most papers that calibrate METANET fail to notice this. List reasons why.
 %            \item We empirically characterize the issue of instability in METANET calibration, specially calibration with gradient-based methods. Then, we propose a rolling-horizon calibration that dynamically assigns METANET parameters and show that this method improves both accuracy of the sim and robustness to perturbation on real data from I-24 MOTION.
 %        \end{itemize}

        Traffic simulation is foundational to the design, assessment, and management of Intelligent Transportation Systems (ITS). Among the available modeling paradigms, macroscopic simulations---which aggregate traffic behavior into continuous fluid dynamics rather than tracking individual vehicles---are particularly valuable due to their fast computation times and scalability. Various studies rely on macroscopic models for traffic state estimation and control \cite{papageorgiou2019role}, with METANET emerging as one of the most popular and performant models~\cite{kotsialos2002traffic, mohammadian2021performance}. 

        The validity of any traffic macrosimulation is predicated upon its effective calibration. Gradient-based optimization methods are desirable for this task. Compared to gradient-free alternatives, gradient-based methods can scale much more efficiently when dealing with the high-dimensional parameter spaces inherent to complex traffic networks~\cite{bouhlel2019gradient, tay2022bayesian}. However, standard gradient-based calibration often struggles when applied to real-world scenarios with multi-segment networks and noisy data~\cite{spiliopoulou2017macroscopic}. The parameter landscapes of macroscopic models like METANET in these settings are highly nonlinear and extremely nonconvex. As a result, gradient-based approaches frequently converge to local minima, yielding parameter sets that produce unstable and unrealistic traffic dynamics \cite{frejo2012parameter}. 

        As a result of the challenges posed by the optimization landscape, we show that a calibrated METANET may exhibit a severe sensitivity to variations in boundary conditions. Even tiny perturbations or system noise can rapidly destabilize the simulated traffic state or generate nonphysical behaviors. Under standard static calibration, where parameters are fixed over time, these small disturbances can compound, compromising the simulation's utility for counterfactual scenario testing and undermining confidence in the estimated parameters. While preliminary observations of this instability have been noted in the literature (e.g., in \cite{mohammadian2021performance}), many studies calibrating METANET do not explicitly account for this sensitivity. This tendency often arises naturally from the standard practice of treating calibration as an isolated, one-time estimation task. By focusing on fitting models to historical training datasets, static approaches can inadvertently bypass the need for out-of-sample stress testing against dynamic, real-world noise.

        To address these limitations, this paper proposes shifting from static parameter estimation to a dynamic approach. This method is motivated by the observation that traffic model parameters, such as capacity, free flow speed, and headway, are not in fact fixed over time, but in reality, they vary with exogenous features such as weather, sun exposure, and time of day   \cite{hranac2006empirical}. Literature in traffic state estimation has also shown that methods that adapt to external, time-varying conditions perform better than a fixed estimator \cite{wang2008real}. We extend this to macrosimulation calibration, where instead of a one-time optimization problem, we treat it as a dynamic calibration problem where parameters vary over time and solve it using a rolling-horizon optimization framework to prevent myopic parameter selections. Beyond allowing for a more expressive model, this dynamic formulation also improves the robustness of the calibrated macrosimulation since each calibration step is solved over a shorter horizon, potentially allowing the solver to search an easier and smoother landscape. We show that by dynamically assigning METANET parameters that change over time, the simulation is able to better replicate traffic phenomena of interest and do so under nontrivial amounts of measurement noise.

        This paper makes three main contributions:
        \begin{itemize}
            \item \textbf{Empirical characterization of instability:} We empirically characterize the instability and sensitivity to perturbations inherent in standard, gradient-based calibration of the METANET macroscopic model.
            \item \textbf{Methodological reformulation:} We reformulate macroscopic model calibration from a static, fixed-parameter estimation task into a dynamic control problem and solve this using a rolling-horizon approach.
            \item \textbf{Real-world validation:} Using real-world trajectory data from the I-24 Mobility Technology Interstate Observation Network (MOTION) testbed, we demonstrate that the proposed dynamic calibration method simultaneously enhances model robustness to perturbations and achieves a 48\% improvement in predictive accuracy over conventional static calibration.
        \end{itemize}
	
	%%%%%%%%%%%%%%%%%%%%%%%%%%%%%%%%%%%%%%%%%%%%%%%%%%%%%%%%%%%%%%%%%%
	\section{Background}
	\label{sec:background}
        \subsection{METANET}
        \label{subsec:background_metanet}
        METANET was proposed in \cite{messmer1990metanet} and is both popular and performant. The original paper has over 700 citations. The benchmarking work in \cite{mohammadian2021performance} found METANET to substantially outperform the six other macroscopic traffic models evaluated.
        
        It is a discrete-time, link-based model, meaning the highway section of interest is split into links of equal length $L$ and the macroscopic traffic variables---density ($\rho$), velocity ($v$), and flow ($q$)---are tracked for each link $\ell \in \mathcal{L}$ at each time step $t \in \mathcal{T}$. The explicit dynamics equations of METANET are as follows:
        \begin{align}
        \rho_{t+1}(\ell) &= \rho_{t}(\ell) + \frac{\delta}{L \lambda_\ell} \left(q_{t}(\ell-1) - \frac{q_{t}(\ell)}{1 - \beta_{\ell}} + r_t \right)
        \label{density}, \\
        v_{t+1}(\ell) &= v_{t}(\ell) + \frac{\delta}{\tau} (V[\rho_{t}(\ell)] - v_{t}(\ell))  \\
        &  \quad \quad + \frac{\delta}{L} \,  v_t(\ell) \left( v_{t}(\ell-1)- v_{t}(\ell)\right) \notag \\
        & \quad \quad - \frac{\nu \delta}{\tau L} \frac{\rho_{t}(\ell+1) - \rho_{t}(\ell)}{\rho_{t}(\ell) + \kappa} \notag, \\
         q_{t}(\ell) &= \rho_{t}(\ell) v_{t}(\ell) \lambda_\ell \label{flow}, \\
        V[\rho_{t}(\ell)] &=  v_{free} \exp \left[-\left(\frac{1}{a} \frac{\rho_{t}(\ell)}{\rho_{cr}}\right)^{a}\right] \label{V},
        \end{align}
        with all METANET parameters not defined above described in Table 1.
        \begin{table}[h!]
            \centering
            \caption{Parameter Definitions for METANET}
            \label{tab:parameters}
            \begin{tabularx}{0.95\linewidth}{>{\raggedright}p{1cm} X}
                \toprule
                \textbf{Parameter} & \text{($^*$ indicates parameter is subject to calibration)} \\
                \midrule
                $\delta$ & Simulation time step in seconds\\
                $L$ & Segment length of each cell in km \\
                $\tau^*$ & Relaxation time constant \\
                $\eta^*$ & Anticipation coefficient \\
                $\kappa^*$ & Numerical stability constant \\
                $a^*$ & Exponential shape of fundamental diagram \\
                $v_{free}^*$& Free flow speed of vehicles \\
                $\rho_{cr}^*$& Critical density of the highway ($\rho > \rho_{cr}$ triggers congestion) \\
                % $\rho_{max}$& Maximum per-lane density of the highway \\
                % $Q$& Flow capacity per lane \\
                $\lambda_{\ell}$ & Number lanes in segment $\ell$  \\
                $\beta_{\ell}^*$ & Turning ratio if an off-ramp exists at link $\ell$\\
                $r_{\ell}^*$ & On-ramp inflow if an on-ramp exists at link $\ell$\\
                \bottomrule
            \end{tabularx}
        \end{table}

        Parameters such as $\delta$ and $L$ are not subject to calibration and are selected to satisfy the Courant-Freidrichs-Lewy (CFL) Condition: $\frac{L}{\delta} \geq v_{\text{free}}$. $\delta$ and $L$ are also contingent on the fidelity of data available and desired granularity of simulation. Values for network-based parameters $\lambda_{\ell}$, $\beta_{\ell}$, and $r_{\ell}$ are typically retrieved from network files (for lane counts and ramp locations) and ramp detector data (for turning ratios and on-ramp inflows). However, in this work, we assume that ground truth ramp data is unavailable or unreliable, which is often the case \cite{muralidharan2009imputation}. Therefore, we allow for the calibration of $\beta$ and $r$ along with the other 6 parameters.

        \subsection{Standard macrosimulation calibration}
        
        Traditionally, macroscopic traffic simulation calibration is treated as an offline, nonlinear optimization problem. The goal is to find a single, optimal set of parameters that minimizes the discrepancy between simulated outputs and historical field data over a fixed time horizon $H$. While theoretically this may be done with gradient-based or gradient-free methods, in practice for multi-segment models with real-world data, gradient-based methods have been found to struggle and gradient-free methods have been preferred~\cite{spiliopoulou2017macroscopic, zhao2025bounded}.
        
        In this formulation, we seek to minimize a loss function $J_{\theta}$:
        \begin{equation}
            \min_{\theta \in \Theta} J_{\theta} = \sum_{t \in \mathcal{T}} \sum_{l \in \mathcal{L}} \left\| x_t^{obs}(\ell) - x_t(\ell; \theta) \right\|^2, \label{calib_start}
        \end{equation} 
        subject to
        \begin{equation}
        \begin{aligned}
            &x_{t+1} = f_{\theta}(x_t, d_t) + w_t, \quad \forall t \in [0, H-1], \\
            &x_0 = x_0^{obs}, \text{and}\\
            &\theta_{min} \leq \theta \leq \theta_{max}.
        \end{aligned}  \label{calib_end}
        \end{equation}
        
         In the above, $\theta$ is a vector of parameters to be calibrated (here, the METANET parameters) from the feasible set $\Theta$, which may represent physical constraints; $\theta_{min}$ and $\theta_{max}$ are the specified lower and upper bounds, respectively, of $\theta$; $x^{obs}_t(\ell)$ is the observed traffic state at segment $\ell$ and time $t$ and $x_t(\ell; \theta)$ is the one produced from a $\theta$-parameterized simulation; $d_t$ models boundary conditions (upstream inflows, downstream conditions) at time $t$ and are typically retrieved from the ground truth data; $w_t$ represents process noise and model mismatch (unmodeled disturbances, driver behavior variability, etc.); and $f_{\theta}(x_t, d_t)$ is the $\theta$-parameterized highway traffic macrosimulation model (i.e. METANET) such that $x_{t+1} = f_{\theta}(x_t, d_t) + w_t$. Note that, in line with the METANET description in Section~\ref{subsec:background_metanet}, the simulation state at segment $\ell$ at time $t$ can be expressed $x_t = \{ v_{t}(\ell), \rho_{t}(\ell), q_{t}(\ell)\}$.
	
	%%%%%%%%%%%%%%%%%%%%%%%%%%%%%%%%%%%%%%%%%%%%%%%%%%%%%%%%%%%%%%%%%%
	\section{Methodology}
    \label{sec:method}

    \subsection{Dynamic macrosimulation calibration} \label{subsec:dynamicpf}
    
    Instead of treating calibration as a one-time estimation task, we reformulate it as a dynamic control problem that adapts model parameters to maintain stability and accuracy under perturbations. We thus reformulate standard macrosimulation calibration as \textbf{dynamic traffic macrosimulation calibration}. Here we seek to minimize a loss function $J(\theta_{0:H-1})$:
    \begin{equation}
        \min_{\theta_t \in \Theta ~\forall t \in \mathcal{T}} J(\theta_{0:H-1}) = \sum_{t \in \mathcal{T}} \sum_{l \in \mathcal{L}} \left\| x_t^{obs}(\ell) - x_t(\ell; \theta_{0:t}) \right\|^2,
    \end{equation}
    subject to the above constraints and additional specifications
    \begin{equation}
    \begin{aligned}
        &\theta_{t+1} = \theta_t + u(x_t, d_t, \theta_t) \text{ and} \\
        &\theta_{min} \leq \theta_t \leq \theta_{max},
    \end{aligned}
    \end{equation}

    where $\theta_{0:H-1}$ is a collection of $\theta_t$ for $t \in [0, H-1]$, $u(x_t, d_t, \theta_t)$ is a feedback law that, given the state, boundary conditions, and parameterization at time $t$, returns parameter residuals. Additionally, the state transition $f_{\theta_t}(\cdot)$ now has time-dependent parameters. Optionally, one can further constrain the optimization via stability constraints ($\| \theta_t - \theta_{nominal} \| \leq \epsilon$), penalization for large parameter jumps ($\| u(x_t, d_t, \theta_t)\| \leq c$), or requirements that the parameters stay fixed over some time period (as we do below).

    \subsection{Rolling Horizon Optimization}

    Given the dynamic calibration formulation in Section~\ref{subsec:dynamicpf}, we now turn to the question of how to determine a feedback law  $u(x_t, d_t, \theta_t)$. In this work, we use \textbf{Rolling Horizon Optimization (RHO)}, a popular optimization technique, also commonly known as Model Predictive Control, that breaks down a long-horizon optimization into shorter sequential subproblems \cite{glomb2022rolling}. Each subproblem is optimized over a prediction horizon $H_p$ and then the solution found is executed for a control horizon $H_c$, where $H_c < H_p$. This reduces the solver's tendency to find solutions that only improve the objective in the near term.

    In the context of calibration, RHO solves the optimization problem for time steps $[t, t+H_p]$ to find $\theta_k$. Then, it simulates the calibrated METANET $f_{\theta_k}$ with boundary conditions $d_{t:t+H_c}$ for $H_c$ time steps, and the resulting state at time $t+H_c$ becomes the starting state for the next subproblem (i.e. $x^{obs}_{t+H_c}$). The following subproblem will then try to find the optimal $\theta_{k+1}$ that minimizes the objective from $[t + H_c, t+ H_c + H_p]$. In the objective for each subproblem, we add a term that is the normalized difference between $\theta_k$ and $\theta_{k+1}$ to penalize large jumps in parameters. In this formulation, parameters change every $H_c$ time steps, and we later show how one can select an appropriate value of $H_c$. Whereas in the experiments below we adjust parameters every $H_c$ time steps, one could alternatively vary parameters more frequently while still predicting and executing parameter selections over $H_p$ and $H_c$ steps. Reformulating the calibration task in this manner can require greater runtime, as repeated optimization procedures must be executed. However, we found these were not prohibitive for our use case since calibration is primarily done offline.
    
    \section{Experiments}

    \begin{figure}[t!]
    \centering
    \includegraphics[width=\linewidth]{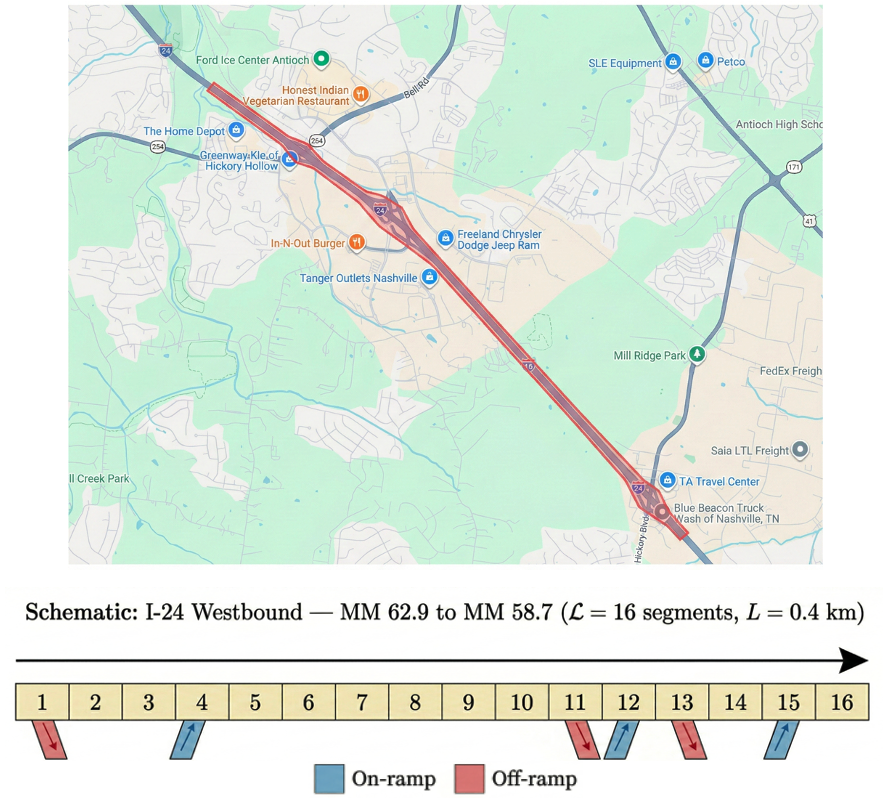}
        \caption{A map view of the stretch of I-24 used in the experiments, highlighted in red (image modified from Google Maps---rights reserved), and a schematic representation of the derived METANET model.}
        \label{fig:i24_network}
    \end{figure}

    To assess the accuracy and robustness of dynamic traffic macrosimulation with RHO, we compare its performance to the standard static approach on a scenario of real-world interest at multiple noise levels. We then explore the behavior of the dynamic approach as a function of the RHO control horizon to understand the balance between over- and under-fitting the observed data; this process also demonstrates how we can select the best RHO control horizon from those assessed in our setting. Finally, to better understand the drivers of robustness, we investigate the local parameter landscapes of the solutions found by the static and dynamic approaches.

    \subsection{Scenario \& Data} 
  
\begin{figure*}[t!]
    \centering
    \includegraphics[width=\textwidth]{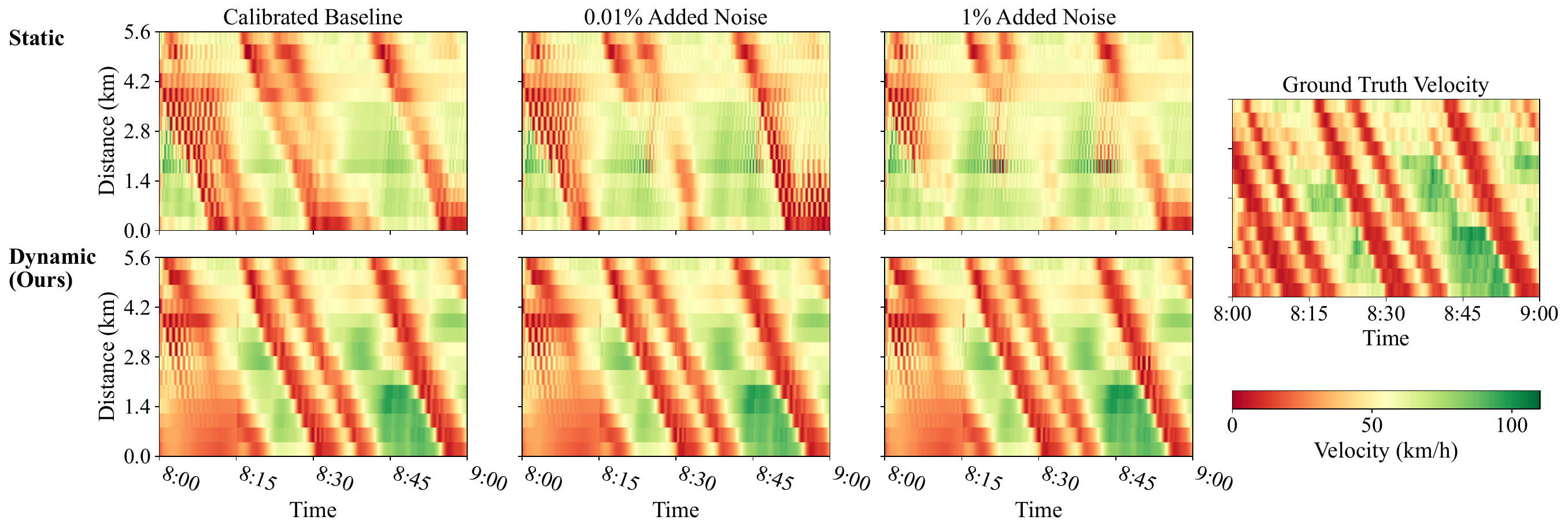}
        \caption{Comparison of speed fields generated by (i) the observed I-24 INCEPTION data used as ground truth (right), (ii) a simulation produced by the standard, static calibration approach (top), and (iii) a simulation produced by the dynamic calibration approach (bottom).}
        \label{fig:tsdiagrams}
    \end{figure*}

   To validate our methodology, we use westbound traffic data from the I-24 MOTION INCEPTION v1.0.0 dataset for 8-9AM on Wednesday, Nov. 30th, 2022. The data is collected via 276 cameras on poles spaced out over 4 miles of the I-24 interstate in Tennessee, USA \cite{gloudemans202324}. See Figure~\ref{fig:i24_network}. The video footage is then converted to vehicle trajectory data. We aggregate the trajectory data into the macroscopic quantities using Edie's defintion and match the granularity needed for our defined METANET model, which is $\delta = 10$ seconds and $L = 0.4$ km. The ground truth speed field is shown in Figure \ref{fig:tsdiagrams}. The spatial segmentation produces a network of 16 segments, with the first and last segments used for boundary conditions. We calibrate parameters for the interior 14 segments, amounting to 5.6km of highway in total. Boundary conditions were smoothed to account for sensor error.

   We select this data partly due to the high-fidelity nature of the trajectories, but mainly because of the traffic phenomenon captured on I-24. Notably, the selected data is during rush hour and shows clear and oscillatory stop-and-go traffic along the entire 4 mile stretch, which provides a scenario that is both challenging for standard calibration methods and of considerable interest. Stop-and-go waves are studied because they are important (contributing to longer travel times, increased emissions, and heightened accident risk), prevalent, and preventable~\cite{kreidieh2018dissipating, he2025review}.

      \begin{figure}[H]
    \centering
    \includegraphics[width=\linewidth]{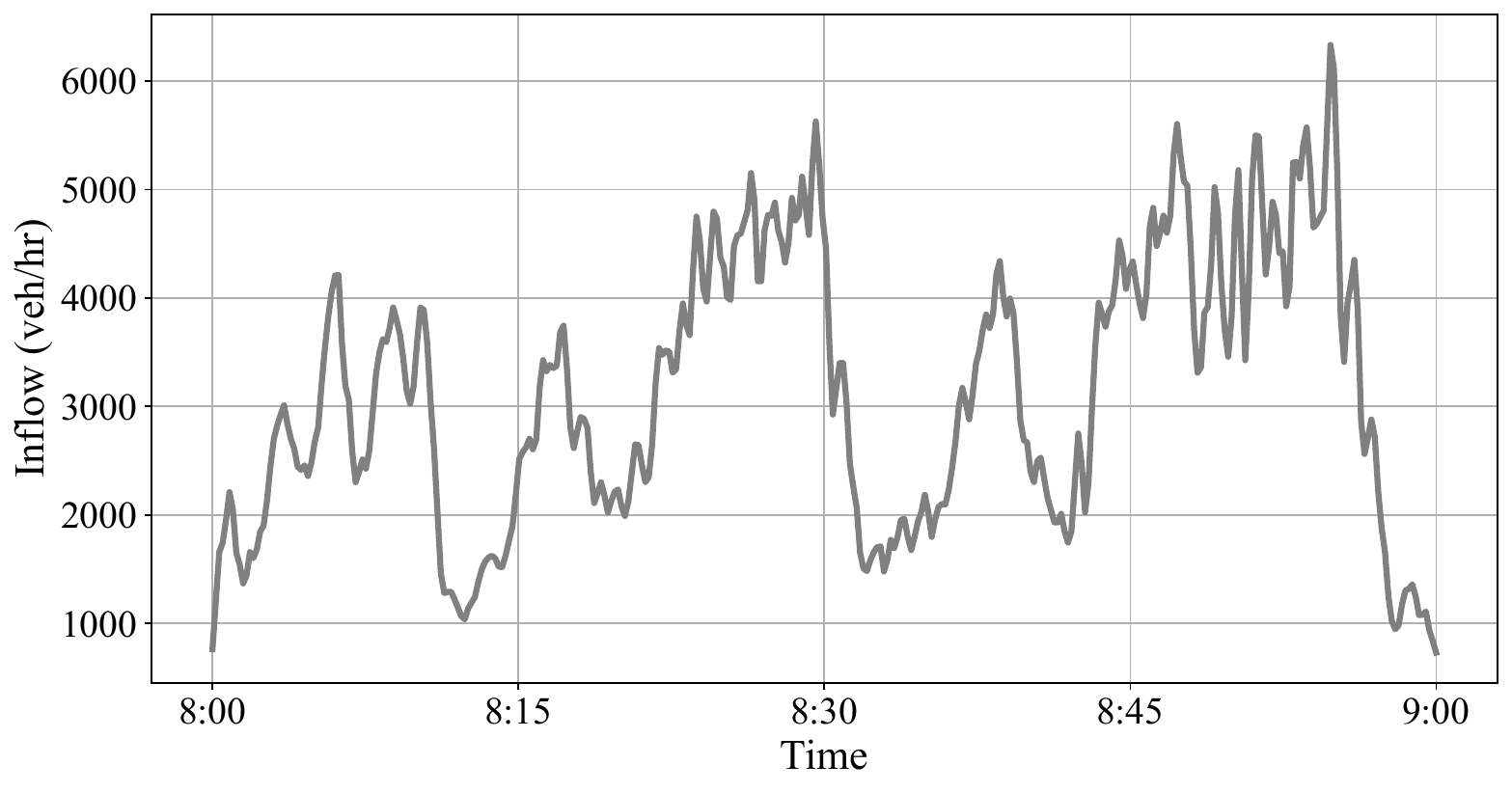}
        \caption{Upstream inflows at the entrance of the I-24 section of interest.}
        \label{fig:inflows}
    \end{figure}
   \subsection{Calibration Specifications}
       For the gradient-based solver in calibration, IPOPT 3.14.16 was used due to its popularity and performance on large-scale nonlinear optimizations problems \cite{biegler2009large}. The calibration framework was implemented in Python 3.12 using the Pyomo optimzation package. The solver was warm-started with METANET parameters from a synthetic network and ran until it found an optimal solution at a tolerance level of $10^{-12}$ \cite{chavoshi2023feedback}. This was executed on an Apple MacBook Pro equipped with an M3 Pro chip and 18GB of unified memory. The code and data for this work is available at \href{https://github.com/srags/dyn-traffic-macrosim}{https://github.com/srags/dyn-traffic-macrosim}.

       For both the baseline (static) and dynamic calibrations, a different set of parameters (and therefore, a different fundamental diagram) was calibrated for each of the 14 segments, since ground truth data was available for every segment. To accord with the known locations of the network's on- and off-ramps, on-ramp inflows were calibrated for segments 4, 12, and 15, and off-ramp turning ratios were calibrated for segments 11 and 13. These were fixed at zero elsewhere. When using dynamic calibration, in addition to being calibrated per segment, parameters were also calibrated over time. The duration over which these were set was specified via the control horizon. Dynamic calibration was run for 7 different control horizons: $[1, 2, 5, 10, 15, 20, 30]$ minutes. The corresponding prediction horizons for each control horizon were selected as $[2, 4, 10, 15, 20, 25, 40]$ minutes.

       In addition to the static gradient-based method, we also compared the dynamic approach to a static gradient-free method. Specifically, we used a Genetic Algorithm (GA), which has been widely applied to macroscopic traffic model calibration due to its ability to handle nonconvex optimization landscapes \cite{spiliopoulou2017macroscopic, poole2012metanet}. GA hyperparameters, including population size, crossover rate, and mutation rate, were selected from \cite{poole2012metanet} and the fitness function was identical to the objective function of the gradient-based methods (see Equation \ref{calib_start}) to ensure a fair comparison across all three approaches.

    \subsection{Evaluation}
    
    To evaluate a calibration method, we assess two quantities: accuracy and robustness. Accuracy is measured using Mean Average Percent Error (MAPE) against the ground truth speed field, which is commonly used in calibration literature \cite{zhao2021methodology}. 
    
    To investigate the robustness of calibrated METANET models, we apply mean-zero Gaussian noise to the ground truth upstream inflows of a calibrated model and run the resulting simulation while keeping the downstream boundary conditions and initial traffic state (at time = 0) fixed. To account for stochasticity, we do this 1,000 times at each noise level for each method. The standard deviation on the Gaussian distribution used for noise application spans several orders of magnitude, from $10^{-7}$\% of each inflow's value and up to 5\%. 
    
    The original (unnoised) flows are plotted in Figure~\ref{fig:inflows}. While the noise applied spans several orders of magnitude, we would expect a realistic and stable simulation to be robust to these perturbations for two reasons. First, note the inflows generally range from 1,000-6,000 vehicles per hour; for reference, 0.01\% of these values amounts to a standard deviation of less than one vehicle per hour (0.1-0.6 vehicles). Second, since the noise is mean-zero, what is applied in one time step is often partially `canceled out' in the following step(s). In other words, the net inflow across the entire 1-hour scenario is, in expectation, the same. 

    \section{Results}

    \subsection{Control Horizon Sensitivity Analysis}

\begin{figure}[]
    \centering
    \includegraphics[width=\linewidth]{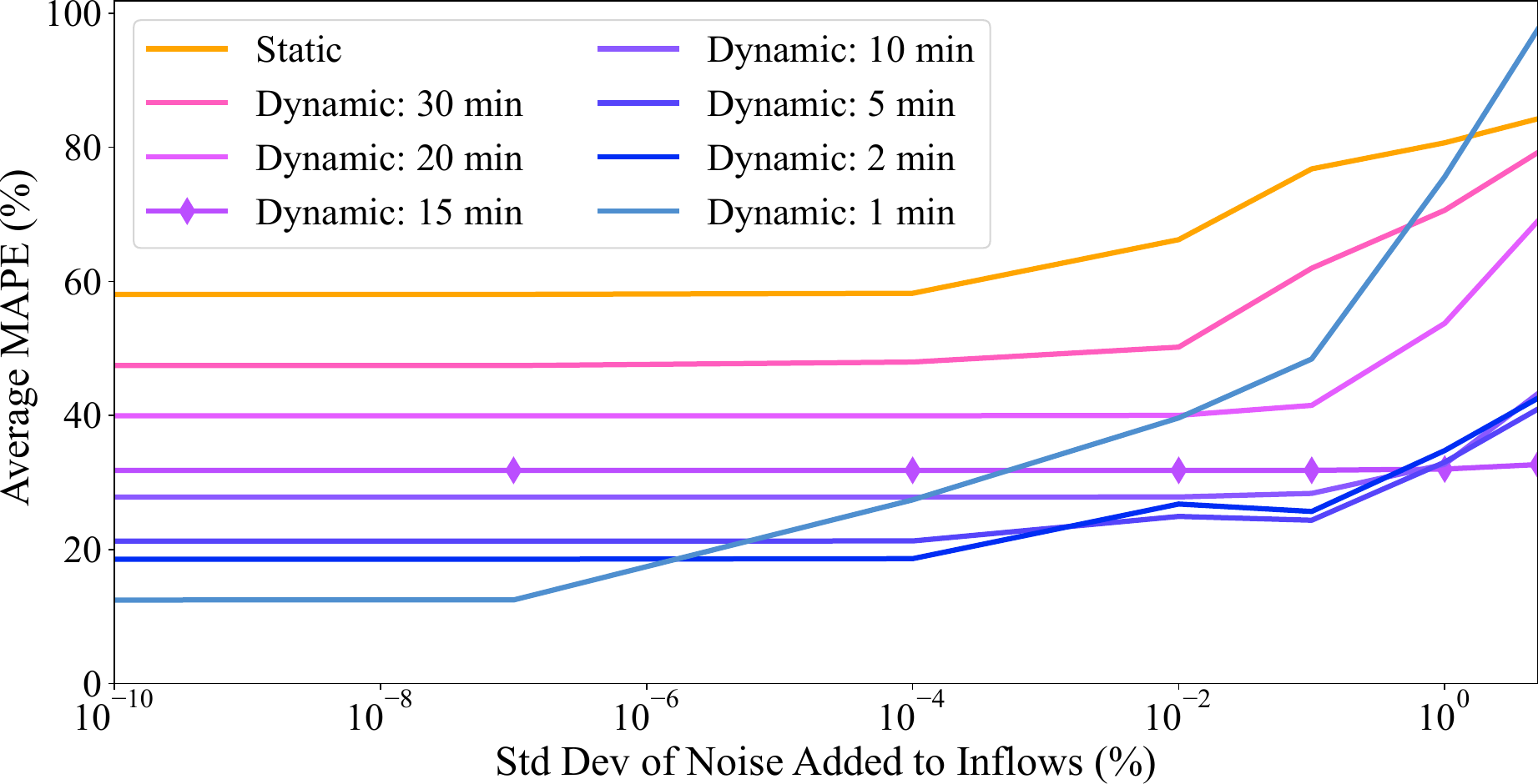}
        \caption{Average MAPE for the static and dynamic (RHO) calibration method across several orders of magnitude of applied mean-zero inflow noise. Dynamic results are shown for various control horizons; the most robust is indicated with diamond  markers.}
        \label{fig:perf_avg}
    \end{figure}

\begin{figure}[]
    \centering
    \includegraphics[width=\linewidth]{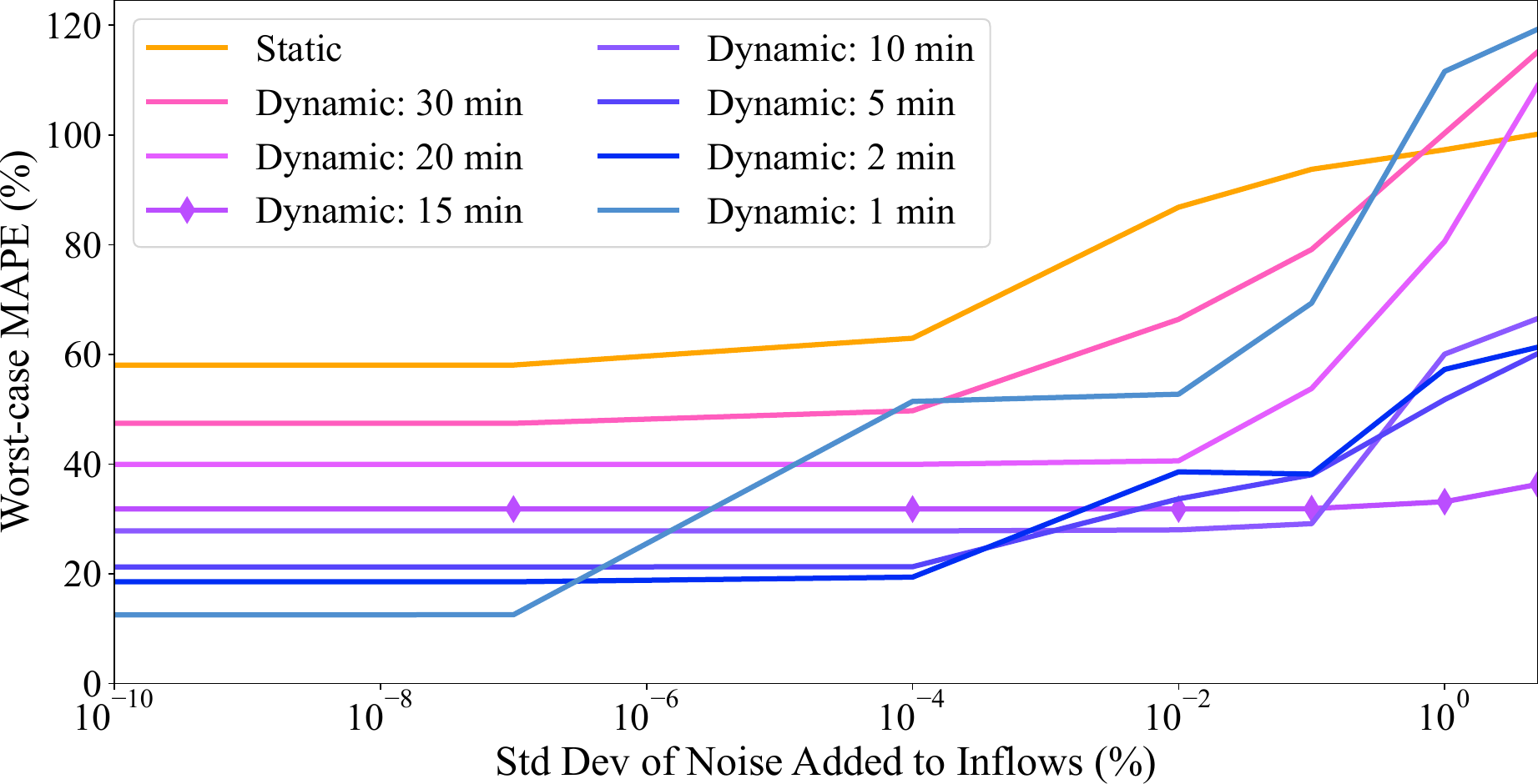}
        \caption{Worst-case MAPE for the static and dynamic (RHO) calibration method across several orders of magnitude of applied mean-zero inflow noise. As in Figure~\ref{fig:perf_avg}, dynamic results are shown for various control horizons and the most robust is indicated with diamond  markers.}
        \label{fig:perf_max}
    \end{figure}

    First, we compare static and rolling horizon optimization of dynamic calibration for various control horizons. In Figures \ref{fig:perf_avg} and \ref{fig:perf_max}, the average and worst-case MAPE at different levels of noised inflows are plotted for the static and all dynamic methods. We see that at both low and high inflow noise, dynamic parameters are able to outperform the standard method, showing that they are generally both more robust and more accurate than standard calibration. In particular, changing the model parameters every 15 minutes is \textbf{48\% more accurate} in the noise-free case, while also \textbf{maintaining its predictive accuracy up to 5\% inflow noise}, well beyond the levels at which the other curves inflect upward. When looking at the worst-case MAPE across 1,000 perturbed inflows in Figure~\ref{fig:perf_max}, relative to Figure~\ref{fig:perf_avg}, the shorter and longer control horizons exhibit worse performance, highlighting the relatively low variance of the 15-minute horizon.

\begin{figure}[t]
    \centering
    \includegraphics[width=\linewidth]{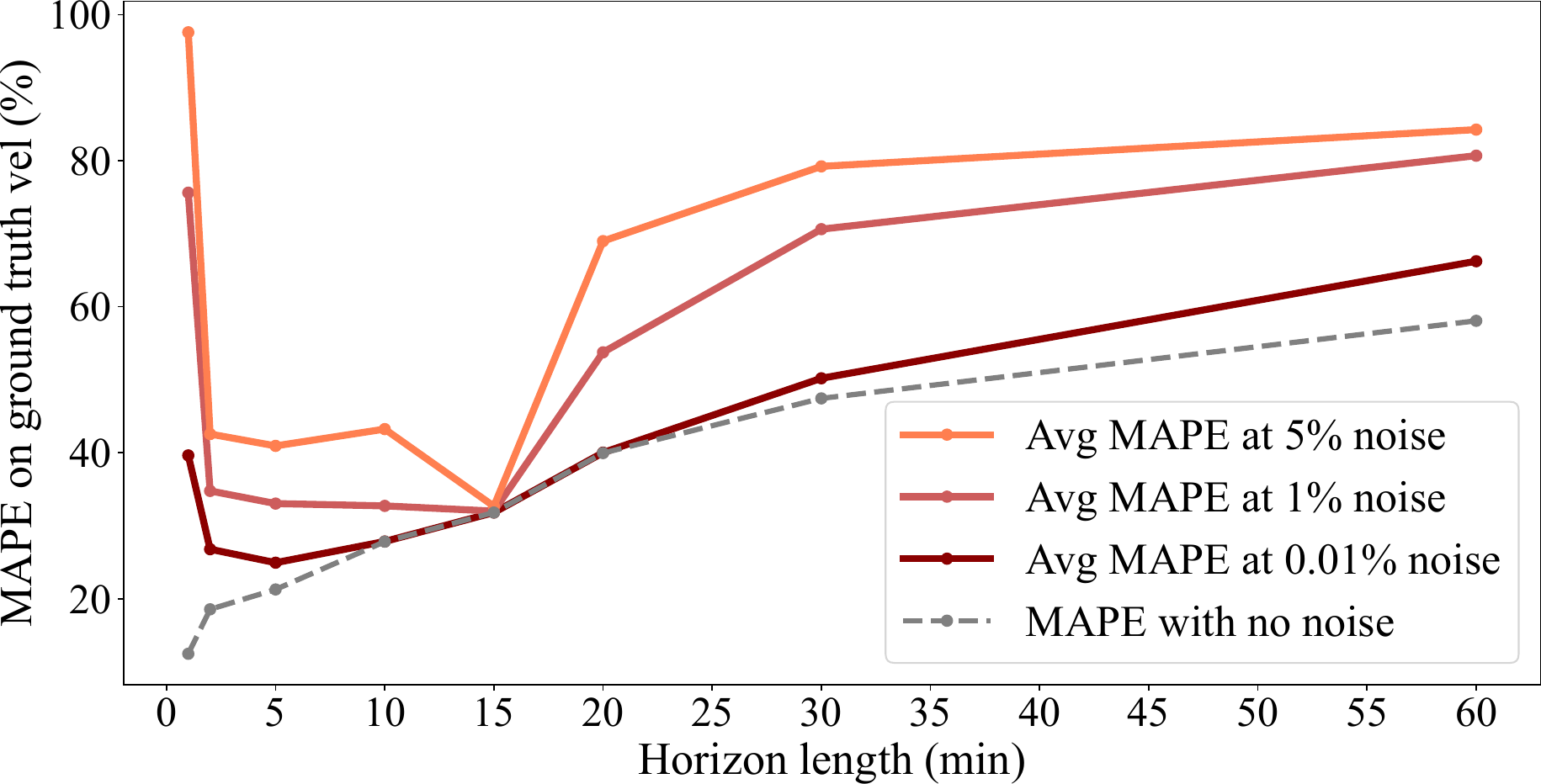}
        \caption{Average MAPE at various noise levels as a function of the dynamic (RHO) method's MPC control horizon $H_c$.}
        \label{fig:dippy_plot}
    \end{figure}

    Figure \ref{fig:dippy_plot} shows the robustness and accuracy tradeoff for different control horizons. The far left of the plot suggests short horizons can overfit the data, resulting in very low error on the unnoised case but failing catastrophically as noise grows. The right side suggests longer horizons underfit: it appears they are not expressive enough to find a valid set of parameters. The dip in the center of the plot shows the point at which the control horizon is able to maintain reasonable accuracy across multiple levels of error. This plot allows us to select an appropriate horizon. For the remaining results, we compare the static calibration to the 15-minute horizon dynamic calibration. This corresponds to the diamond-marked lines in Figures~\ref{fig:perf_avg} and~\ref{fig:perf_max}.

    \subsection{Static vs. Dynamic Comparison}

    \begin{figure}[]
    \centering
    \includegraphics[width=\linewidth]{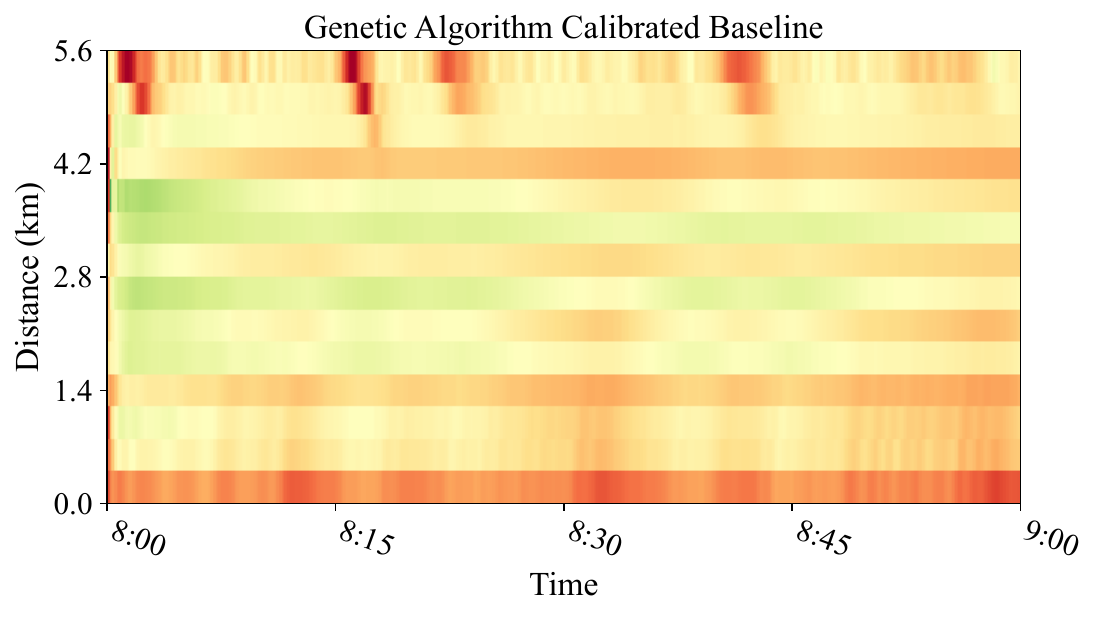}
        \caption{Speed field generated by the (static) gradient-free, GA-calibrated METANET model on the unperturbed I-24 INCEPTION dataset. It fails to reproduce the ground truth stop-and-go waves seen in Figure \ref{fig:tsdiagrams}.}
        \label{fig:ga_ts}
    \end{figure}
    
      \begin{figure}[]
    \centering
    \includegraphics[width=0.9\linewidth]{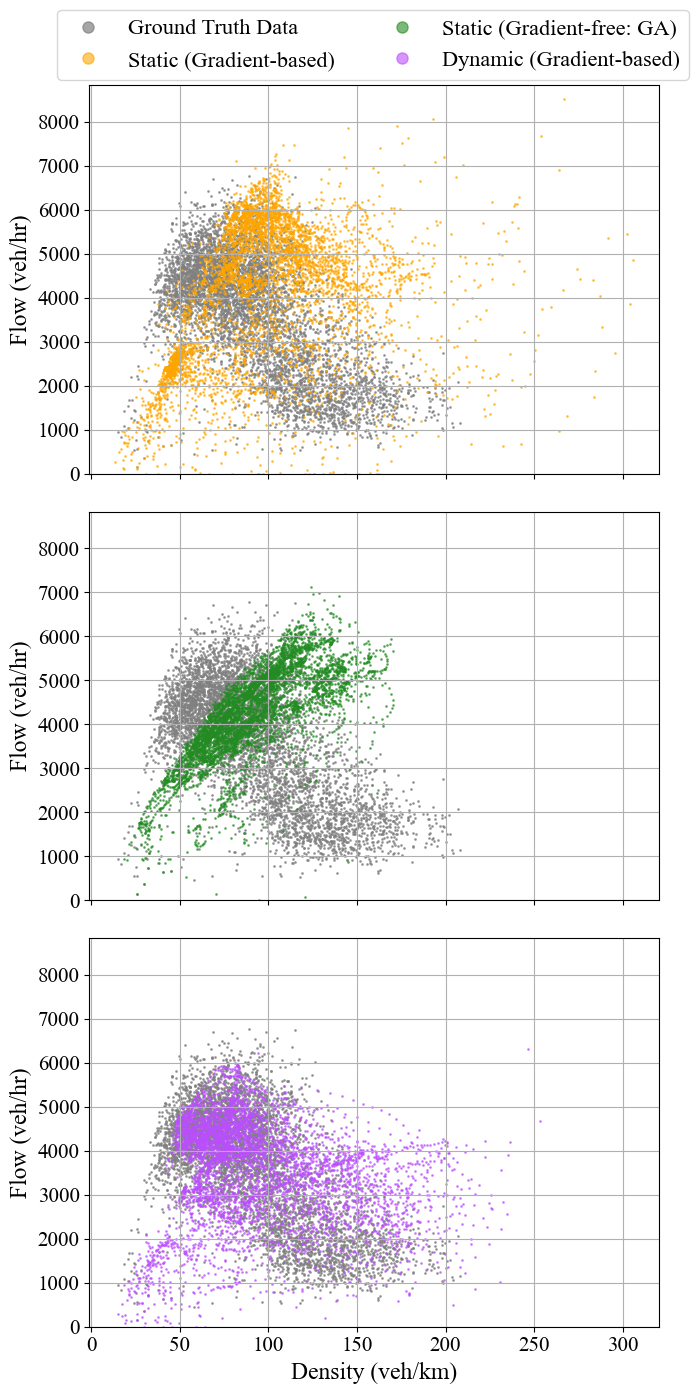}
        \caption{Fundamental diagrams showing flow and density values across all times and I-24 calibrated segments in the unperturbed setting.}
        \label{fig:FD_comparison}
    \end{figure}

    The performance gaps between the static and the best-performing dynamic method are shown in Figures~\ref{fig:tsdiagrams}, \ref{fig:ga_ts}, and \ref{fig:FD_comparison}. Figure~\ref{fig:tsdiagrams} depicts the robustness gap and its impact on the simulations produced by each. The simulated speed fields with the worst MAPE from sampling 1,000 perturbed inflows is shown for 0.01\% noise and 1\% noise; we observe that the dynamically calibrated model withstands the noise much better than the standard gradient-based calibration. The phenomenon of interest---stop-and-go waves---degrades almost to non-existence with 1\% added noise in the standard case, and it even shows signs of numerical instability around 1.8 km at 8:20AM. Figure~\ref{fig:ga_ts} shows that the static gradient-free method fails to reproduce the stop-and-go waves, even under unperturbed conditions, thus producing behavior inconsistent with the observed data. While the gradient-free GA offers robustness advantages by avoiding the sharp local minima characteristic of gradient-based optimization, robustness to noise is only meaningful if the underlying calibration is accurate to begin with.
    
    To understand the accuracy gap, Figure~\ref{fig:FD_comparison} compares the fundamental diagrams (FD) of the static gradient-based, static GA, and dynamic calibrated models in the unnoised setting. It highlights that the dynamic parameters provide a simulation that is a better fit to the ground truth data compared to both static baselines. The points overlap more in the dynamic FD, especially at the critical density, and the dynamically calibrated model has fewer erroneous high flow and density values. The GA-calibrated model produces a cluster of points more misaligned with the fundamental diagram and struggles to represent both the free-flow and congested regime, providing evidence in further support of gradient-based methods.

    \subsection{Parameter Landscape Analysis}
        \begin{figure*}[t]
    \centering
    \includegraphics[width=\textwidth]{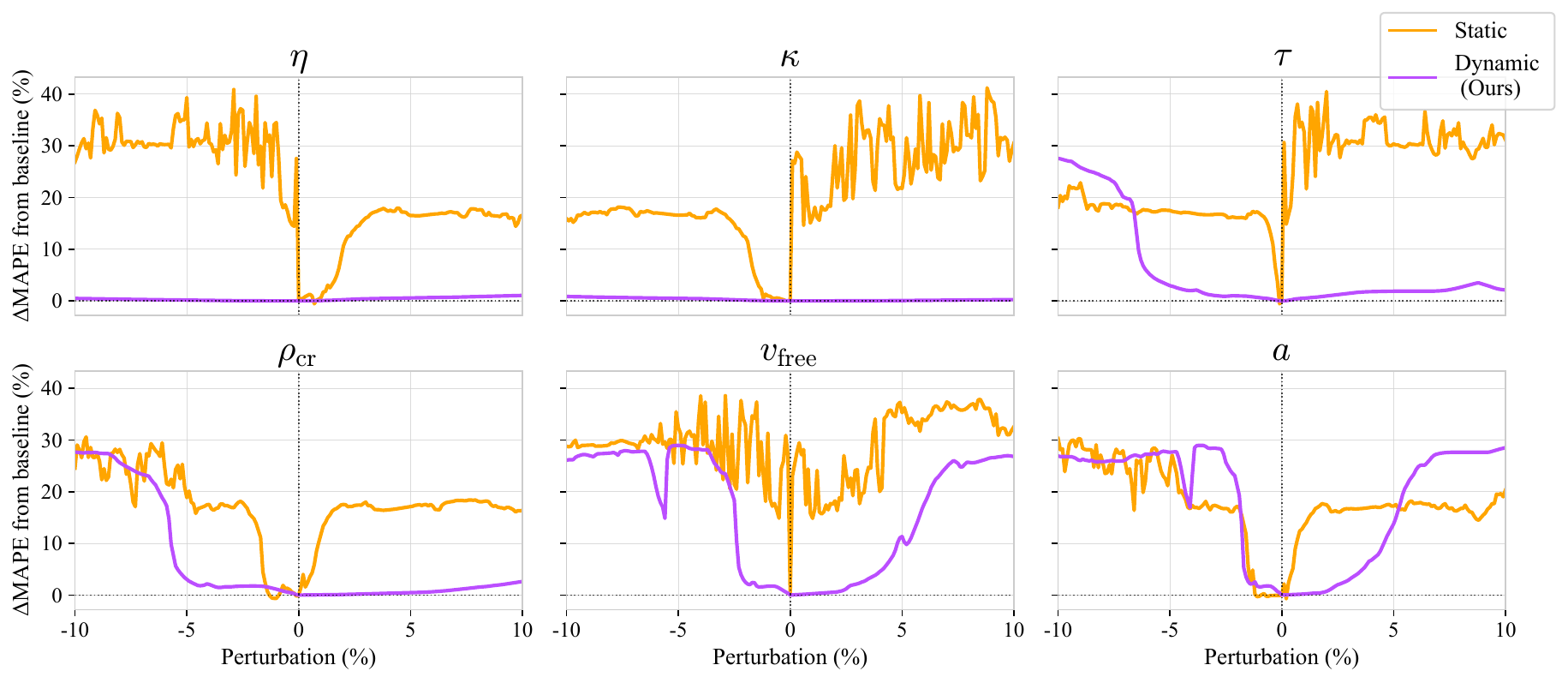}
        \caption{Comparison of the parameter-error landscapes for the standard and dynamic calibration processes. Note that the y-axis is the MAPE difference from their respective baselines, so all landscapes start at 0\% $\Delta$MAPE when there is no perturbation.}
        \label{fig:sensitivity}
    \end{figure*}
    To better understand the space over which the solver is optimizing and whether, as hypothesized, the dynamic method allows the system to find smoother and less sharp (local) optima, we plot perturbed parameter landscapes for the gradient-based methods. See Figure~\ref{fig:sensitivity}. For both the static and best dynamic parameters, each of the six main METANET parameters are perturbed on the first calibrated segment from -10\% to +10\% of their post-calibration value for all time steps. Ramp flows are excluded since this segment is ramp-free. The parameters for the remaining segments remain fixed to their calibrated values. The model is simulated with the perturbed parameter, and the speed MAPE difference from the baseline (i.e. no perturbation) is plotted. The landscapes demonstrate that dynamic parameters are less sensitive to parameter perturbations and the optima found by the dynamic method are in wider and smoother basins than those found via the static approach. This further affirms the robust behavior we observed for the RHO method.

    \section{Conclusion}
    
        Accurate calibration of macroscopic traffic models like METANET is essential, yet standard gradient-based methods struggle with highly nonlinear, nonconvex parameter landscapes. Consequently, these static approaches frequently produce unrealistic traffic dynamics and severe sensitivity to minor boundary perturbations. To overcome this, we introduced a dynamic, rolling-horizon calibration framework. By reformulating static parameter estimation into a dynamic control problem, our method adapts parameters over time to maintain stability and accuracy amid disturbances.
        
        Experiments using real-world data validate this approach. The framework improves predictive accuracy by 48\% over conventional static calibration in unperturbed settings and achieves a better fit to the ground truth data. Unlike the static, gradient-based method, our framework also demonstrates robustness to varying levels of measurement noise. Furthermore, landscape analysis revealed that our dynamic method guides the solver toward wider, smoother optima, successfully avoiding the brittle minima characteristic of standard techniques. 
        
        Future research should conduct analytical investigations of the observed instabilities, explore adapting the dynamic framework to other macroscopic simulation models, and test it across broader data distributions to ensure robustness consistency. While this study considered freeway corridors, future work could also evaluate the framework's generalizability to urban networks by adapting the approach to models suited for arterial topologies and signalized intersections. Finally, investigating its sensitivity to parameter bounding constraints and initial conditions could help refine the calibration process.

%%%%%%%%%%%%%%%%%%%%%%%%%%%%%%%%%%%%%%%%%%%%%%%%%%%%%%%%%%%%%%%%%%
	\section*{ACKNOWLEDGMENTS}
	% Replace with acknowledgments or remove if none
	This work was done with the support of the NSF Graduate Research Fellowship under Grant No. 2141064.
%%%%%%%%%%%%%%%%%%%%%%%%%%%%%%%%%%%%%%%%%%%%%%%%%%%%%%%%%%%%%%%%%%
	%\addtolength{\textheight}{-12cm}
	%\vspace{10mm}
	\bibliographystyle{IEEEtran}
	% Your .bib file here
	\bibliography{root} 

@article{bouhlel2019gradient,
  title={Gradient-enhanced kriging for high-dimensional problems},
  author={Bouhlel, Mohamed A and Martins, Joaquim RRA},
  journal={Engineering with Computers},
  volume={35},
  number={1},
  pages={157--173},
  year={2019},
  publisher={Springer}
}

@article{he2025review,
  title={A Review of Stop-and-Go Traffic Wave Suppression Strategies: Variable Speed Limit vs. Jam-Absorption Driving},
  author={He, Zhengbing and Laval, Jorge and Han, Yu and Hegyi, Andreas and Nishi, Ryosuke and Wu, Cathy},
  journal={arXiv preprint arXiv:2504.11372},
  year={2025}
}

@article{kotsialos2002traffic,
  title={Traffic flow modeling of large-scale motorway networks using the macroscopic modeling tool METANET},
  author={Kotsialos, Apostolos and Papageorgiou, Markos and Diakaki, Christina and Pavlis, Yannis and Middelham, Frans},
  journal={IEEE Transactions on intelligent transportation systems},
  volume={3},
  number={4},
  pages={282--292},
  year={2002},
  publisher={IEEE}
}

@inproceedings{kreidieh2018dissipating,
  title={Dissipating stop-and-go waves in closed and open networks via deep reinforcement learning},
  author={Kreidieh, Abdul Rahman and Wu, Cathy and Bayen, Alexandre M},
  booktitle={2018 21st international conference on intelligent transportation systems (itsc)},
  pages={1475--1480},
  year={2018},
  organization={IEEE}
}

@article{messmer1990metanet,
  title={METANET: A macroscopic simulation program for motorway networks},
  author={Messmer, Albert and Papageorgiou, Markos},
  journal={Traffic engineering \& control},
  volume={31},
  number={9},
  year={1990}
}

@article{mohammadian2021performance,
  title={Performance of continuum models for realworld traffic flows: Comprehensive benchmarking},
  author={Mohammadian, Saeed and Zheng, Zuduo and Haque, Md Mazharul and Bhaskar, Ashish},
  journal={Transportation Research Part B: Methodological},
  volume={147},
  pages={132--167},
  year={2021},
  publisher={Elsevier Ltd.}
}

@incollection{papageorgiou2019role,
  title={The role of macroscopic modeling in the simulation, surveillance and control of motorway network traffic},
  author={Papageorgiou, Markos and Papamichail, Ioannis and Wang, Yibing},
  booktitle={Transport Simulation},
  pages={3--25},
  year={2019},
  publisher={EPFL Press}
}

@article{spiliopoulou2017macroscopic,
  title={Macroscopic traffic flow model calibration using different optimization algorithms},
  author={Spiliopoulou, Anastasia and Papamichail, Ioannis and Papageorgiou, Markos and Tyrinopoulos, Yannis and Chrysoulakis, John},
  journal={Operational Research},
  volume={17},
  number={1},
  pages={145--164},
  year={2017},
  publisher={Springer}
}

@article{tay2022bayesian,
  title={Bayesian optimization techniques for high-dimensional simulation-based transportation problems},
  author={Tay, Timothy and Osorio, Carolina},
  journal={Transportation Research Part B: Methodological},
  volume={164},
  pages={210--243},
  year={2022},
  publisher={Elsevier}
}

@article{zhao2025bounded,
  title={Bounded-METANET: A new discrete-time second-order macroscopic traffic flow model for bounded speed},
  author={Zhao, Weiming and Roncoli, Claudio and Yildirimoglu, Mehmet},
  journal={Transportation Research Part C: Emerging Technologies},
  volume={180},
  pages={105345},
  year={2025},
  publisher={Elsevier}
}

@inproceedings{frejo2012parameter,
  title={A parameter identification algorithm for the METANET model with a limited number of loop detectors},
  author={Frejo, Jos{\'e} Ram{\'o}n Dom{\'\i}nguez and Camacho, Eduardo Fern{\'a}ndez and Horowitz, Roberto},
  booktitle={2012 IEEE 51st IEEE Conference on Decision and Control (CDC)},
  pages={6983--6988},
  year={2012},
  organization={IEEE}
}

@article{hranac2006empirical,
  title={Empirical studies on traffic flow in inclement weather},
  author={Hranac, Rob and Sterzin, Emily D and Krechmer, Daniel and Rakha, Hesham and Farzaneh, Mohamadreza},
  year={2006}
}

@article{wang2008real,
  title={Real-time freeway traffic state estimation based on extended Kalman filter: Adaptive capabilities and real data testing},
  author={Wang, Yibing and Papageorgiou, Markos and Messmer, Albert},
  journal={Transportation Research Part A: Policy and Practice},
  volume={42},
  number={10},
  pages={1340--1358},
  year={2008},
  publisher={Elsevier}
}

@article{muralidharan2009imputation,
  title={Imputation of ramp flow data for freeway traffic simulation},
  author={Muralidharan, Ajith and Horowitz, Roberto},
  journal={Transportation Research Record},
  volume={2099},
  number={1},
  pages={58--64},
  year={2009},
  publisher={SAGE Publications Sage CA: Los Angeles, CA}
}

@article{glomb2022rolling,
  title={A rolling-horizon approach for multi-period optimization},
  author={Glomb, Lukas and Liers, Frauke and R{\"o}sel, Florian},
  journal={European Journal of Operational Research},
  volume={300},
  number={1},
  pages={189--206},
  year={2022},
  publisher={Elsevier}
}

@article{gloudemans202324,
  title={I-24 MOTION: An instrument for freeway traffic science},
  author={Gloudemans, Derek and Wang, Yanbing and Ji, Junyi and Zachar, Gergely and Barbour, William and Hall, Eric and Cebelak, Meredith and Smith, Lee and Work, Daniel B},
  journal={Transportation Research Part C: Emerging Technologies},
  volume={155},
  pages={104311},
  year={2023},
  publisher={Elsevier}
}

@article{biegler2009large,
  title={Large-scale nonlinear programming using IPOPT: An integrating framework for enterprise-wide dynamic optimization},
  author={Biegler, Lorenz T and Zavala, Victor M},
  journal={Computers \& Chemical Engineering},
  volume={33},
  number={3},
  pages={575--582},
  year={2009},
  publisher={Elsevier}
}

@article{chavoshi2023feedback,
  title={A feedback linearization approach for coordinated traffic flow management in highway systems},
  author={Chavoshi, Kimia and Ferrara, Antonella and Kouvelas, Anastasios},
  journal={Control Engineering Practice},
  volume={139},
  pages={105615},
  year={2023},
  publisher={Elsevier}
}

@article{zhao2021methodology,
  title={Methodology for Calibration and Validation of Mesoscopic Traffic Simulation Models},
  author={Zhao, MO and Appiah, Justice},
  year={2021}
}

@article{poole2012metanet,
  title={METANET model validation using a genetic algorithm},
  author={Poole, A and Kotsialos, A},
  journal={IFAC Proceedings Volumes},
  volume={45},
  number={24},
  pages={7--12},
  year={2012},
  publisher={Elsevier}
}
	
\end{document}